\documentclass[aps,prl,preprint]{revtex4}

\usepackage{amsmath,amssymb,epsfig}
\usepackage{graphicx}

\usepackage[T1]{fontenc}

\begin{document}

\title{\bf Electronic structure of FeSe monolayer superconductors}

\author{I.A. Nekrasov$^a$, N.S. Pavlov$^a$, M.V. Sadovskii$^{a,b}$,
A.A. Slobodchikov$^a$
\\
{$^a$Institute for Electrophysics, Russian Academy of Sciences, Ural Branch}\\
{\sl Amundsen str. 106, Ekaterinburg, 620016, Russia}\\
{$^b$M.N. Mikheev Institute for Metal Physics, Russian Academy of Sciences,
Ural Branch}\\ {\sl S. Kovalevsky str. 18, Ekaterinburg, 620290, Russia}}


\begin{abstract}

We review a variety of theoretical and experimental results concerning electronic
band structure of superconducting materials based on FeSe monolayers.
Three type of systems are analyzed: intercalated FeSe systems
A$_x$Fe$_2$Se$_{2-x}$S$_x$ and [Li$_{1-x}$Fe$_x$OH]FeSe as well as the single
FeSe layer films on SrTiO$_3$ substrate.
We present the results of detailed first principle electronic band structure
calculations for these systems together with comparison with some experimental
ARPES data. The electronic structure of these systems is rather different from
that of typical FeAs superconductors, which is quite significant for possible
microscopic mechanism of superconductivity. This is reflected in the absence
of hole pockets of the Fermi surface at $\Gamma$-point in Brillouin zone, so
that there are no ``nesting'' properties of different Fermi surface pockets.
LDA+DMFT calculations show that correlation effects on Fe-3d states in the
single FeSe layer are not that strong as in most of FeAs systems. As a result,
at present there is no theoretical understanding of the formation of rather
``shallow'' electronic bands at $M$ points. LDA calculations show that the
main difference in electronic structure of FeSe monolayer on SrTiO$_3$
substrate from isolated FeSe layer is the presence of the band of O-2p surface
states of TiO$_2$ layer on the Fermi level together with Fe-3d states, which may
be important for understanding the enhanced $T_c$ values in this system.
We briefly discuss the implications of our results for microscopic models of
superconductivity.


\end{abstract}

\maketitle




\section{Introduction}

The discovery of a new class of superconductors based on iron pnictides  has
opened up the new prospects for the study of high-temperature superconductivity
(cf. reviews \cite{Sad_08, Hoso_09, John, MazKor, Stew, Kord_12}).
The nature of superconductivity in these novel materials and other physical
properties significantly differs from those of high -- $T_c$ cuprates, though
they still have many common features, which gives hope for a better
understanding of the problem of high-temperature superconductivity in general.

The discovery of superconductivity in iron pnictides was soon followed by its
discovery in iron {\em chalcogenide} FeSe. A lot of attention was attracted to
this system because of its simplicity, though its superconducting characteristics
(under normal conditions) were quite modest ($T_c\sim$8K) and its electronic
structure was quite similar to that of iron pnictides (cf. review in \cite{FeSe}).

The situation with iron chalcogenides fundamentally changed with the appearance
of {\em intercalated} FeSe based systems with the value of $T_c\sim$30-40K,
which immediately attracted attention due to their unusual electronic structure
\cite{JMMM, JTLRev}. Currently quite the number of such compounds is known.
The first systems of this kind were A$_x$Fe$_{2-y}$Se$_2$ (A=K,Rb,Cs) with the
value of $T_c\sim$ 30K \cite{AFeSe, AFeSe2}. It is generally believed that
superconductivity in this system appears in an ideal 122-type structure.
However samples studied so far always have been multiphase, consisting of a
mixture of mesoscopic superconducting and insulating (antiferromagnetic)
structures such as K$_2$Fe$_4$Se$_5$, which complicates the studies of this
system.

A substantial further increase of $T_c$ up to 45K has been achieved by
intercalation of FeSe layers with rather large molecules in compounds such as
Li$_x$(C$_2$H$_8$N$_2$)Fe$_{2-y}$Se$_2$ \cite{intCH} and
Li$_x$(NH$_2$)$_y$(NH$_3$)$_{1-y}$Fe$_2$Se$_2$ \cite{intNH}.
The growth of $T_c$ in these systems might be associated with increase of the
distance between the FeSe layers from 5.5\AA~ to $\sim$7\AA~
in A$_x$Fe$_{2-y}$Se$_2$ and 8-11\AA~in the large molecules intercalated systems,
i.e. with the growth of the two-dimensional character of the materials.
Most recently the active studies has started of [Li$_{1-x}$Fe$_x$OH]FeSe system
with the value of $T_c\sim$43K \cite{LiOH1, LiOH2}, where a good enough
single -- phase samples and single crystals were obtained.

A significant breakthrough in the study of iron based superconductors happened
with the observation of a record high $T_c$ in epitaxial films of single FeSe
monolayer on a substrate of SrTiO$_3 $(STO) \cite{FeSe_STO1}.
These films were grown as described in Ref. \cite{FeSe_STO1} and most of the
works to follow on the 001 plane of the STO.
The tunnel experiments reported in Ref. \cite{FeSe_STO1} produced the record
values of the energy gap, while the resistivity measurements gave the temperature
of the beginning of superconducting transition substantially higher than 50K.
It should be noted that the films under study are very unstable on the air.
Thus in most works the resistive transitions were mainly studied on films
covered with amorphous Si or several FeTe layers. It significantly reduces the
observed values of $T_c$. Unique measurements of FeSe films on STO, done in Ref.
\cite{FeSe_STO2} {\em in situ}, gave the record value of $T_c$>100K. So far,
these results have not been confirmed by the other authors. However ARPES
measurements of the temperature dependence of the superconducting gap in such
films, now confidently demonstrate value of $T_c$ in the range of 65--75 K.

Films consisting of several FeSe layers produce the values of $T_c$
significantly lower than those for the single -- layer films \cite{FeSe13UCK}.
Recently  monolayer FeSe film on 110 STO plane \cite{FeSe_STO_FeTe} covered with
several FeTe layers was grown. Resistivity measurements on these films (including
measurements of the upper critical magnetic field $ H_ {c2} $) gave value of
$T_c\sim$30K. At the same time, the FeSe film, grown on BaTiO$_3$ (BTO) substrate,
doped with Nb (with even larger than in STO values of the lattice constant
$\sim$ 3.99 \AA), showed in the ARPES measurements the value of $T_c\sim $ 70K
\cite{FeSe_BTO}. In a recent paper \cite{FeSe_Anatas} it was reported the
observation of quite high values of the superconducting gap in FeSe (from
tunnelling spectroscopy) for FeSe monolayers grown on 001 plane of TiO$_2 $
(anatase), which in its turn was grown on the 001 plane of SrTiO$_3$.
The lattice constant of anatase is very close to the lattice constant of bulk
FeSe, so FeSe film remains essentially unstretched.

Single -- layer FeSe films were also grown on the graphene substrate, but the value
of $T_c$ was of the order of 8-10K similar to bulk FeSe \cite{FeSeGraph}.
That emphasizes the role of the unique properties of substrates such as
Sr(Ba)TiO$_3$, which can play determining role in the significant increase of
$T_c$.

More information about the experiments on single -- layer FeSe films can be
found in the recently published review of Ref. \cite{FeSe1UC_rev}.
Below we shall concentrate on the present day understanding of the electronic
structure of FeSe monolayer systems.

\section{Crystal structures of iron based superconductors}

Bulk FeSe system has the simplest crystal structure among other iron high-T$_c$
superconductors. A unit cell is a tetrahedron with Fe ion in the center and Se
in its vertices. The symmetry group is P4/nmm with lattice constants ---
$a = 3.765$ \AA~(Fe-Fe distance) and $c = 5.518$ \AA~(interlayer distance),
with the height of the Se ions over the Fe planes $z_{Se} = 0.2343$
\AA~\cite{param}.

\begin{figure}
\includegraphics[clip=true,width=0.8\textwidth]{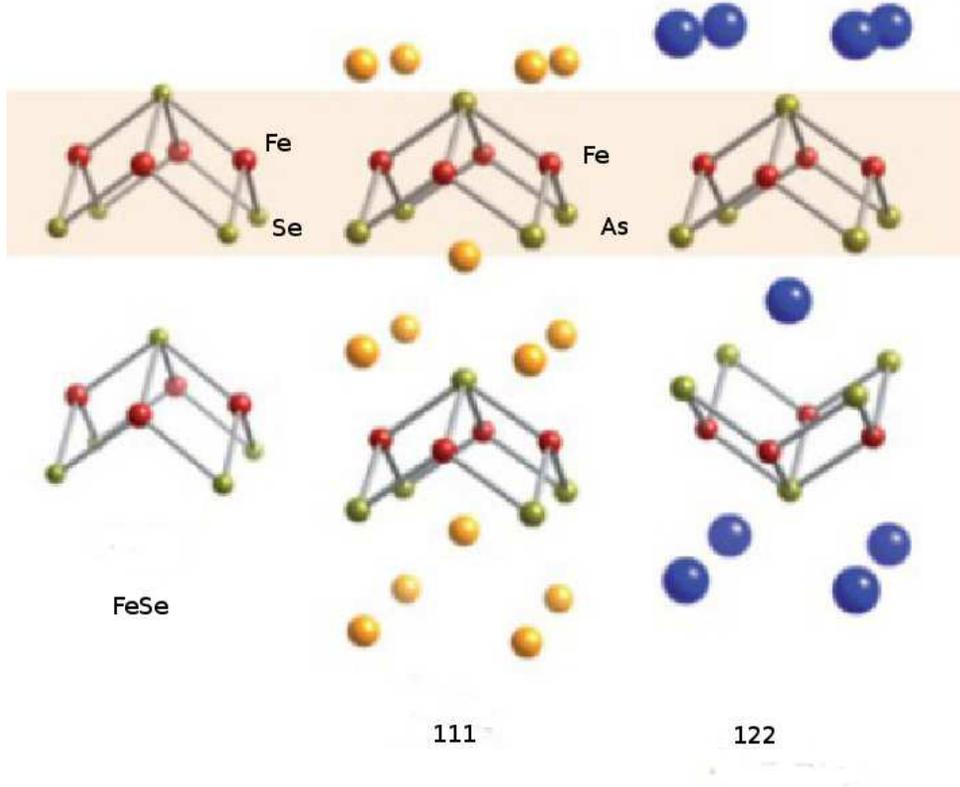}
\caption{Typical crystal structures of iron based superconductors.}
\label{StrucFePn}
\end{figure}

Figure \ref{StrucFePn} schematically shows a simple crystal
structure of iron based superconductors
\cite{ FeSe, Sad_08, Hoso_09, John, MazKor, Stew, Kord_12}. The common element
for all of them is the presence of the FeAs or FeSe plane (layer), where Fe ions
form a simple square lattice. The pnictogen (Pn - As) or chalcogen (Ch - Se) ions
are located at the centers of the squares above and below Fe plane in a staggered
order. The 3d states of Fe in FePn plane (Ch) play a decisive role in the
formation of the electronic properties of these systems, including superconductivity.
In this sense, these layers are quite similar to the CuO$_2$ planes in cuprates
(copper oxides). Also these systems can be considered, to a first approximation,
as a quasi-two dimensional.

\begin{figure}
\includegraphics[clip=true,width=0.4\textwidth]{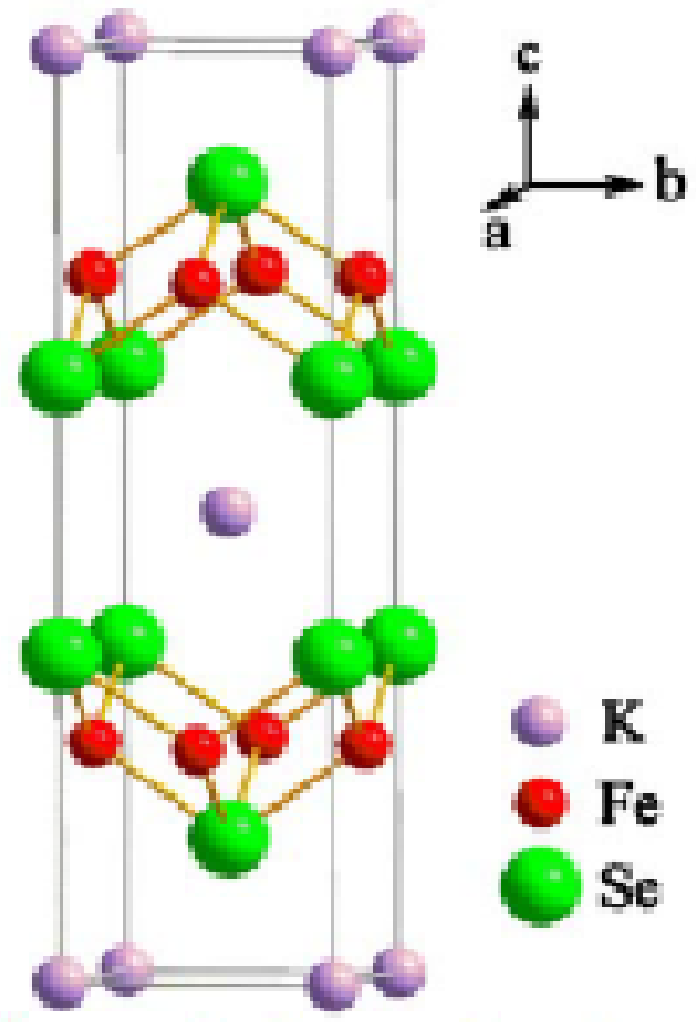}
\includegraphics[clip=true,width=0.35\textwidth]{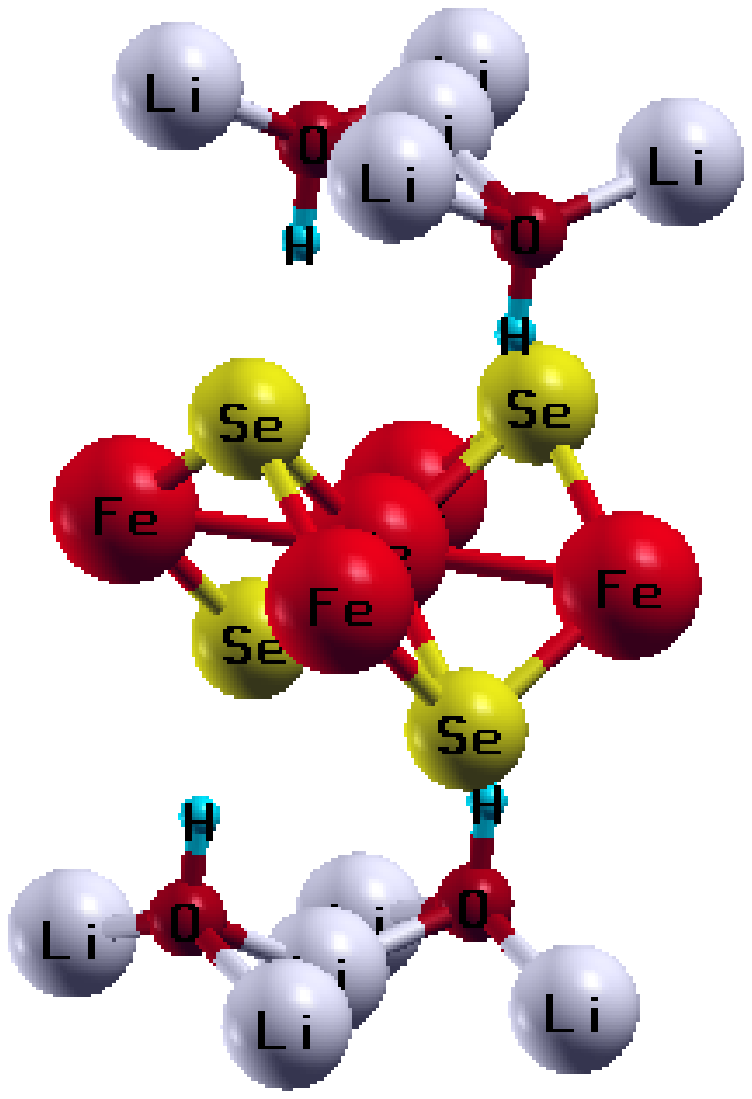}
\caption{(a) --  ideal ($x$=1) crystal structure 
of 122-type of K$_x$Fe$_2$Se$_2$, (b) -- ideal ($x$=0)
crystal structure of [Li$_{1-x}$F$_x$OH]FeSe compound.}
\label{interFeCh}
\end{figure}

Note that all of the FeAs crystal structures shown in Fig. \ref{StrucFePn}
are ion--covalent crystals. Chemical formula, say for a typical system 122, can
be written for example as Ba$^{+2}$(Fe$^{+2}$)$_2$(As$^{-3}$)$_2$. Charged FeAs
layers are held by Coulomb forces from the surrounding ions.
In the bulk FeSe electrically neutral FeSe layers are held by much weaker
van der Waals interactions.
This makes the system suitable for intercalation of various atoms and molecules
that can be fairly easy introduced between the layers of FeSe. Chemistry of
intercalation processes for iron chalcogenide superconductors
is discussed in detail in a recent review Ref. \cite{VivRod}.
The crystal structure of K$_x$Fe$_2$Se$_2$ and [Li$_{1-x}$F$_x$OH]FeSe systems
are shown in Fig. \ref{interFeCh}(b).

\begin{figure}
\includegraphics[clip=true,width=0.8\textwidth]{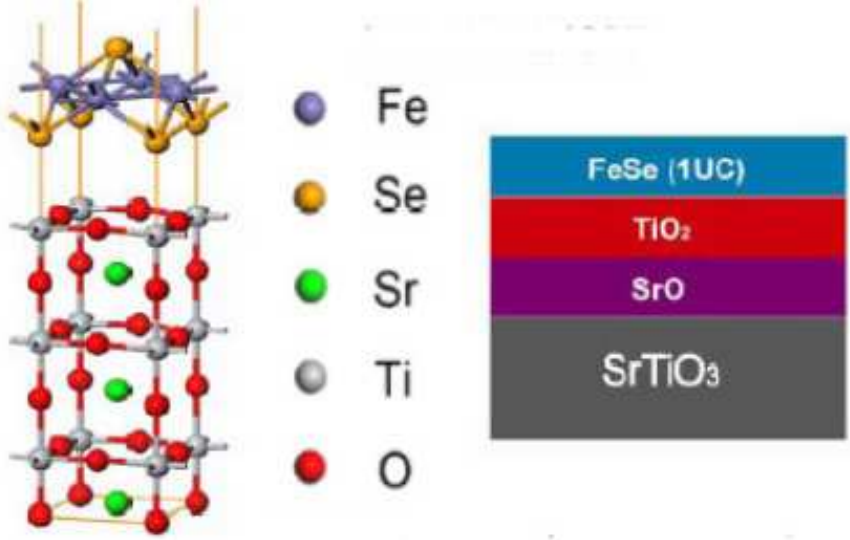}
\caption{Crystal structure of FeSe monolayer on (001) surface of SrTiO$_3$.}
\label{FeSeSTO}
\end{figure}

The structure of the FeSe monolayer film on STO is shown in Fig. \ref{FeSeSTO}.
It can be seen that the FeSe layer is directly adjacent to the surface TiO$_2$
layer of STO. Note that the lattice constant within FeSe layer in a bulk samples
is equal to 3.77\AA, while STO has substantially greater lattice constant equal
to 3.905 \AA. Thus the single -- layer FeSe film should be noticeably stretched,
compared with the bulk FeSe. However this tension quickly disappears as the
number of subsequent layers grows.

\section{Electronic structure of iron -- selenium systems}

Electronic spectrum of iron pnictides now is well understood, both by theoretical
calculations based on the modern band structure theory and ARPES experiments
\cite{Sad_08, Hoso_09, John, MazKor, Stew, Kord_12}.
Almost all physical effects of interest to superconductivity are determined by
electronic states of FeAs plane (layer), shown in Fig. \ref{StrucFePn} (b).
The spectrum of carriers in the vicinity of the Fermi level $\pm$ 0.5 eV, where
superconductivity is formed, practically have only Fe-3$d$ character.
Thus the Fermi level is crossed by four or five bands (two or three hole and two
electronic ones), forming a typical semi -- metallic dispersions.

In this rather narrow energy interval around the Fermi level the dispersions can
be considered as parabolic \cite{MazKor,KuchSad10}. LDA+DMFT calculations
\cite{DMFT1, DMFT2} show that the role of electronic correlations
in iron pnictides, unlike the cuprates, is relatively insignificant. It is
reduced to a noticeable renormalization of the effective masses of the electron
and hole dispersions, as well as to general ``compression'' (reduction) of
the bandwidth.

The presence of the electron and hole Fermi surfaces of similar size, satisfying
(approximately!) the ``nesting'' condition plays a very important role in the
theories of superconducting pairing in iron arsenides based on (antiferromagnetic)
spin fluctuations \cite{MazKor}. We shall see below that the electronic spectrum
and Fermi surfaces in the Fe chalcogenides are very different from this
qualitative picture of Fe pnictides. It raises the new problems for the
explanations of microscopic mechanism of superconductivity in FeSe systems.

\subsection{A$_x$Fe$_2$Se$_2$  system}

\begin{figure}
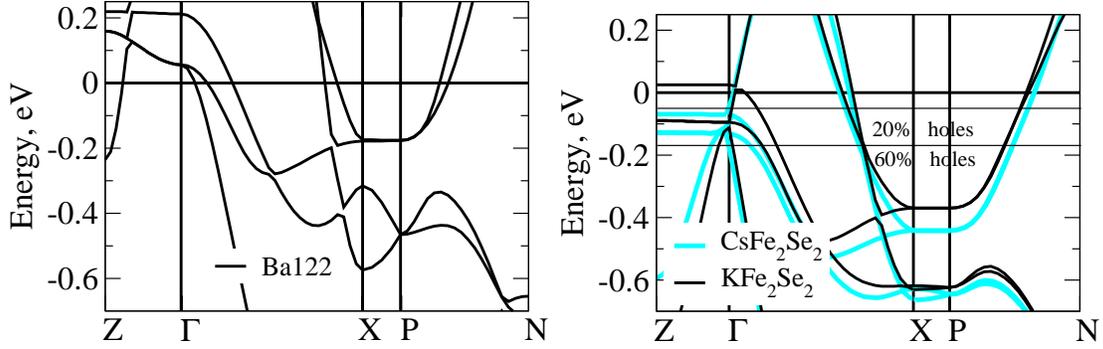

\includegraphics[clip=true,width=0.45\textwidth]{Ba122_Ef_bands.eps}
\includegraphics[clip=true,width=0.45\textwidth]{KCs_Ef_bands.eps}
\caption{(a) -- LDA bands of Ba122 near the Fermi level ($E=$0) \cite{Ba122},
(b) -- LDA bands of K$_x$Fe$_2$Se$_2$ (black lines) and Cs$_x$Fe$_2$Se$_2$
(blue lines). Additional horizontal lines correspond to
Fermi levels of 20\% and 60\% hole doping \cite{KFe2Se2}.} 
\label{122comp}
\end{figure}

LDA calculations of electronic structure of the A$_x$Fe$_{2-y}$Se$_2$ (A=K,Cs)
system were performed immediately after its experimental discovery
\cite{KFe2Se2, KFe2Se2SI}. Surprisingly enough, this spectrum has appeared to be
qualitatively different from that of the bulk FeSe and spectra of all known
systems based on FeAs. In Fig. \ref{122comp} we compare energy bands of
BaFe$_2$As$_2 $ (Ba122) \cite{Ba122} (the typical prototype of FeAs systems)
and A$_x$Fe$_{2-y}$Se$_2$(A=K,Cs) \cite{KFe2Se2}.
One can see a significant difference in the spectra near the Fermi level.

\begin{figure}
\includegraphics[clip=true,width=0.9\textwidth]{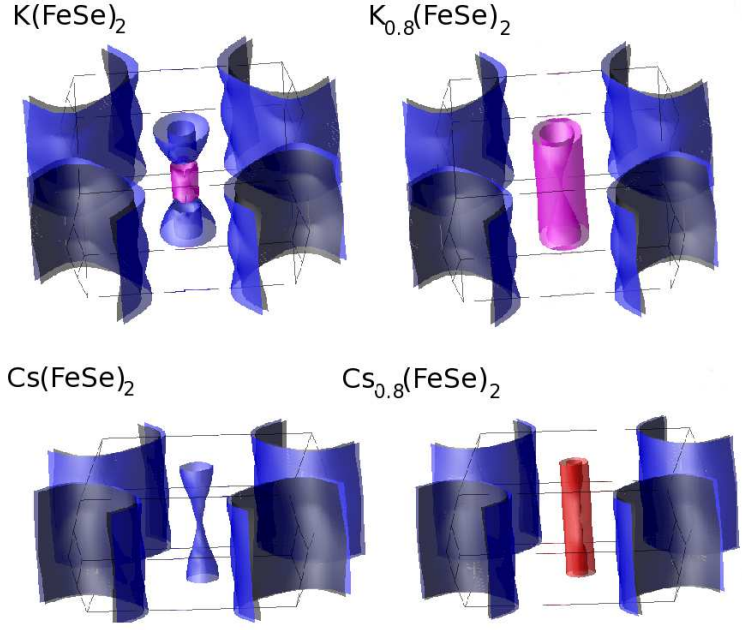}
\caption{LDA Fermi surfaces A$_x$Fe$_2$Se$_2$ (A=K,Cs) for the stoichiometric
(left) and 20\% (right) hole doping \cite{KFe2Se2}. }
\label{FSAFeSe}
\end{figure}

In Fig. \ref{FSAFeSe} we show the calculated  Fermi surfaces for two systems
A$_x$Fe$_{2-y}$Se$_2$ (A=K,Cs) at various doping levels \cite{KFe2Se2}.
We see that they differ significantly from the Fermi surfaces of FeAs systems
Fermi surfaces --- in the center of the Brillouin zone, there are only small
(mainly electronic!) Fermi sheets, while the electronic cylinders in the
Brillouin zone corners are substantially larger.
The shape of the Fermi surface, typical for bulk FeSe and FeAs systems,
can be reproduced only at a much larger (experimentally inaccessible) levels of
the hole doping \cite{KFe2Se2}.

\begin{figure}
\includegraphics[clip=true,width=0.75\textwidth]{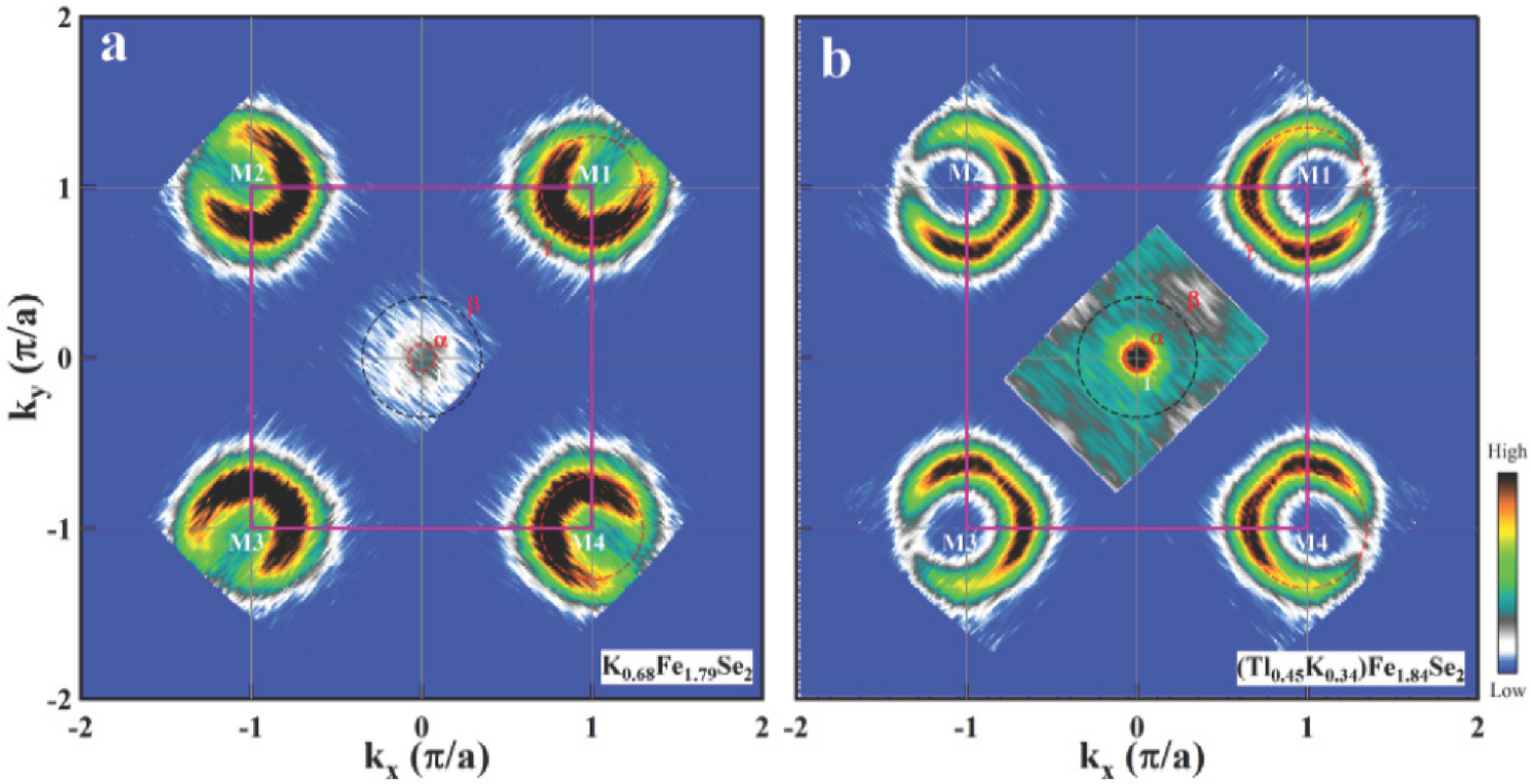}
\caption{ARPES Fermi surfaces 
of K$_{0.68}$Fe$_{1.79}$Se$_2$ 
($T_c$=32K) and  Tl$_{0.45}$K$_{0.34}$Fe$_{1.84}$Se$_2$ ($T_c$=28K) 
\cite{KFe2Se2ARPES}.}   
\label{FSFeSeARP}
\end{figure}

This shape of the Fermi surfaces in A$_x$Fe$_{2-y}$Se$_2$ systems
was rather soon supported by ARPES experiments.
For example, in Fig. \ref{FSFeSeARP} we show ARPES data of Ref.
\cite{KFe2Se2ARPES}, which obviously is in agreement with LDA data of
Refs. \cite{KFe2Se2,KFe2Se2SI}.
One can clearly see that in this system it is impossible to speak of any, even
approximate, ``nesting'' properties of electron and hole Fermi surfaces.

LDA+DMFT calculations for K$_{1-x}$Fe$_{2-y}$Se$_2$ system for various doping
levels were done in  Refs. \cite{KFeSeLDADMFT1,KFeSeLDADMFT2}. There, along with
the standard LDA+DMFT approach, we also used our LDA$'$+DMFT \cite{CLDA, CLDA_long}
approach, which allows, in our opinion, to solve the problem of ``double counting''
of Coulomb interaction in the LDA+DMFT in a more consistent way.
For DMFT calculations Coulomb and exchange interactions of the electrons in the
Fe-3d shell we have chosen $U = 3.75$ ~ eV and $ J = 0.56 $ eV and as
an {\em impurity solver} Hirsh -- Fye Quantum Monte-Carlo algorithm (QMC) was used.
The results of the LDA calculations are useful to compare with the ARPES data
obtained in Refs. \cite{KFe2Se2_ARPES,KFe2Se2_ARPES_2}.

It is turned out that for K$_{1-x}$Fe$_{2-y}$Se$_2$ correlation effects play
quite an important role. They lead to a noticeable change in LDA energy dispersions.
In contrast to iron arsenides, where the quasiparticle bands near the Fermi level
are well defined, in the K$_{1-x}$Fe$_{2-y}$Se$_2$ compounds in the vicinity of
the Fermi level there is a strong suppression of quasiparticle bands.
This reflects the fact that the correlation effects in this system are stronger
than in iron arsenides. The value of the quasiparticle renormalization
(correlation narrowing) of the bands at the Fermi level is 4-5, whereas in iron
arsenides this factor is only 2-3 for the same values of the interaction parameters.

The results of these calculations, in general, are in good qualitative agreement
with the ARPES data \cite{KFe2Se2_ARPES, KFe2Se2_ARPES_2}.
Both demonstrate strong damping of quasiparticles in the immediate vicinity of
the Fermi level and a strong renormalization of the effective mass as compared
to systems based on FeAs. However, in our calculations formation of unusually
``shallow'' (depth $\sim $ 0.05 eV) electron bands near the $X$-point
in the Brillouin zone, observed in ARPES experiments, is not visible.

\begin{figure}[!ht]
\includegraphics[width=.99\textwidth]{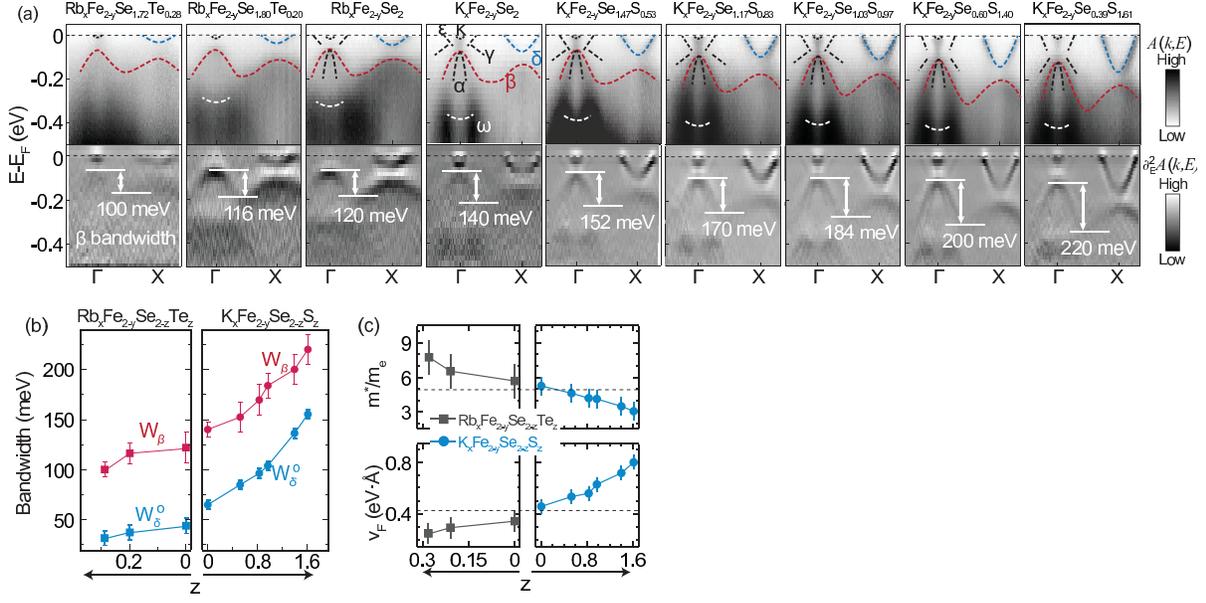}
\caption{ (a) Doping dependence of the ARPES along the $\varGamma$-$X$ 
direction in K$_x$Fe$_{2-y}$Se$_{2-z}$S$_z$. 
(b) The bandwidth of $\delta$ and $\beta$ bands as a function of doping. 
(c) Doping dependence of the effective mass ($m^*$) and Fermi velocity 
($v_F$) of the $\delta$ band \cite{KFe2Se2_ARPES_2}.}
\label{fig1}
\end{figure}

In Ref. \cite{KFe2Se2_ARPES_2} the authors reported systematic ARPES
study of the K$_x$Fe$_{2-y}$Se$_{2-z}$S$_{z}$ system at different doping
levels. It was shown that the sulfur doping level $z$ can control the depth of
the  ``shallow'' electron band  near the $X$-point
(Fig.~\ref{fig1}, $\delta$ -- band). We tried to model this situation
for different compositions of K$_x$Fe$_{2-y}$Se$_{2-z}$S$_{z}$, taking into
account the changes of lattice constant. We considered three cases of
K$_x$Fe$_{2-y}$Se$_{2}$, K$_x$Fe$_{2-y}$Se$_{1}$S$_{1}$ and
K$_x$Fe$_{2-y}$Se$_{0.4}$S$_{1.6}$. The appropriate values of lattice constants
used in our calculations are listed in Table~\ref{Tab1}.

\begin{figure}
\includegraphics[width=.5\textwidth]{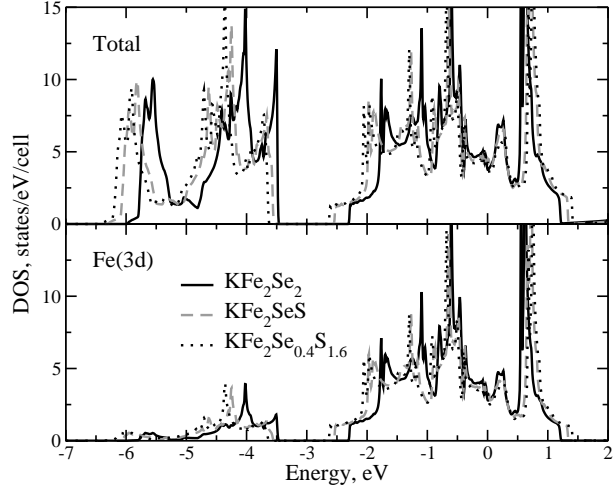}
\caption{LDA density of states of K$_x$Fe$_{2-y}$Se$_{2-z}$S$_{z}$ compounds.}
\label{fig2}
\end{figure}

\begin{figure}
\includegraphics[width=.4\textwidth]{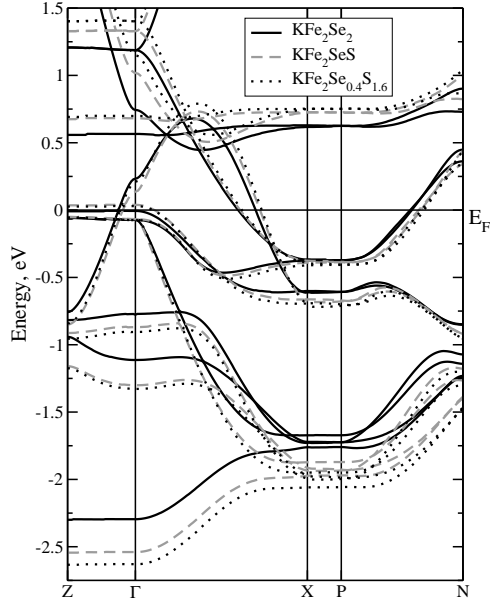}
\caption{LDA band dispersions of K$_x$Fe$_{2-y}$Se$_{2-z}$S$_{z}$ compounds.}
\label{fig3}
\end{figure}

\begin{table*}[tbp]
\caption{Lattice constants
for K$_x$Fe$_{2-y}$Se$_{2-z}$S$_{z}$ \cite{KFe2Se2,KFeSeS_struct_11},
LDA bandwidth of Fe-3d bands (eV) and energy interval of
Se-4p states.}
\begin{tabular}{lcccc}
\hline\hline
K$_{x}$Fe$_{2-y}$Se$_{2-z}$S$_{z}$ & a (\AA ) & c (\AA )
& Bandwidth of Fe-3d states (eV) & Selenium states energies (eV) \\
\hline
KFe$_{2}$Se$_{2}$                         & 3.9136 & 14.0367  &  3.5  & (-5.8;-3.45) \\
K$_{0.70}$Fe$_{1.55}$Se$_{1.01}$S$_{0.99}$ & 3.805 & 13.903   &  3.9  & (-6.1;-3.55) \\
K$_{0.80}$Fe$_{1.64}$Se$_{0.42}$S$_{1.58}$ & 3.781 & 13.707   &  4.0  & (-6.2;-3.6)\\
\hline\hline
\end{tabular}
\label{Tab1}
\end{table*}

The LDA calculated density of states and band dispersions are presented in
Fig.~\ref{fig2} and Fig.~\ref{fig3} correspondingly. Upon sulfur doping
the bandwidth of Fe-3d states increases, while the bottom of the $\delta$ -- band
slightly goes down. Corresponding $\delta$ -- band bottom positions are
0.37, 0.38, 0.40 eV. The Se-4p states also go down in energy
(see Table~\ref{Tab1}). These results are in qualitative agreement with ARPES
experiments, though the unusually low values of band bottom energies
(``shallow'' band formation) of $\delta$ -- band remain a mystery. We can also
note, that according to Table~\ref{Tab1} K$_2$Fe$_2$Se$_2$ has the smallest
(LDA) bandwidth of 3d -- states and it grows with sulfur doping. Thus one can
expect, that K$_2$Fe$_2$Se$_2$ is the most correlated system in this series.

\subsection{[Li$_{1-x}$Fe$_x$OH]FeSe system}

\begin{figure}
\includegraphics[clip=true,width=0.35\textwidth]{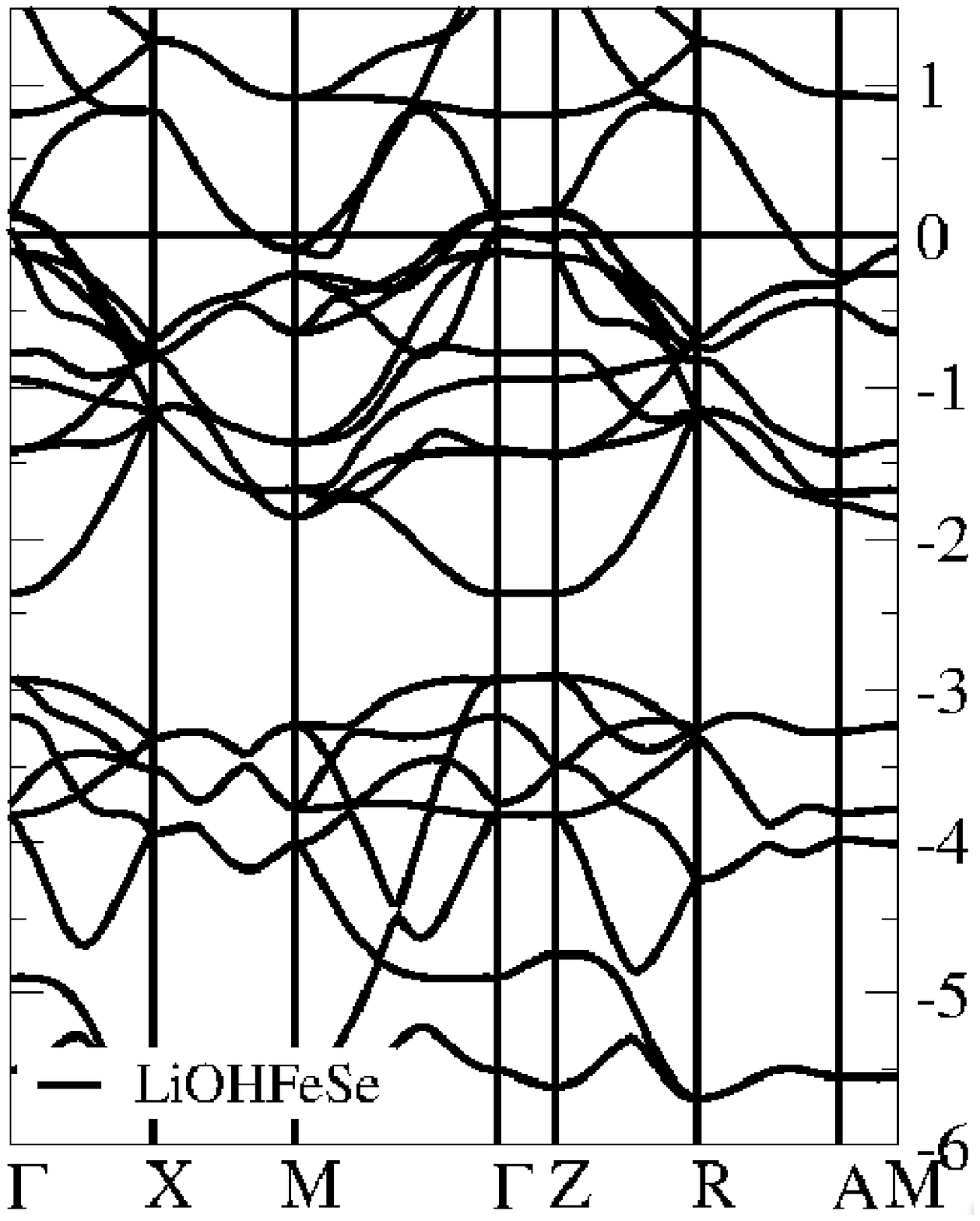}
\includegraphics[clip=true,width=0.40\textwidth]{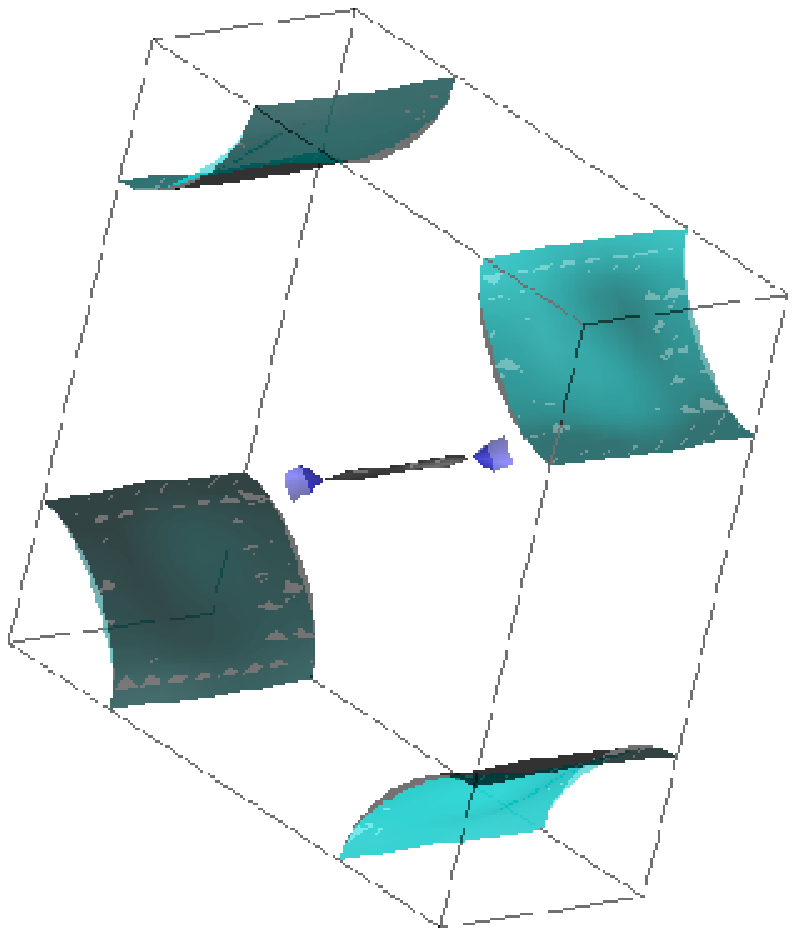}
\includegraphics[clip=true,width=0.45\textwidth]{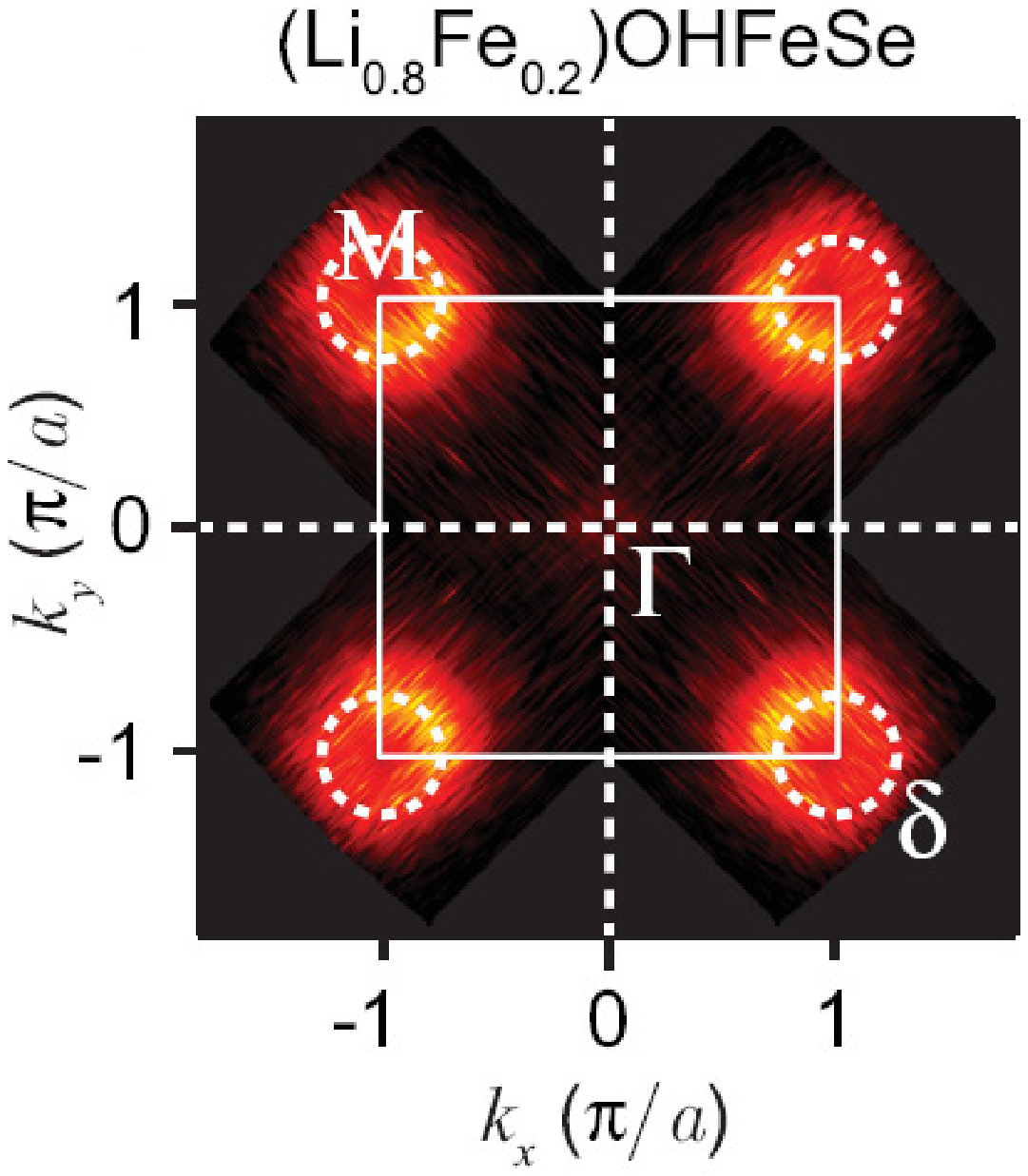}
\includegraphics[clip=true,width=0.45\textwidth]{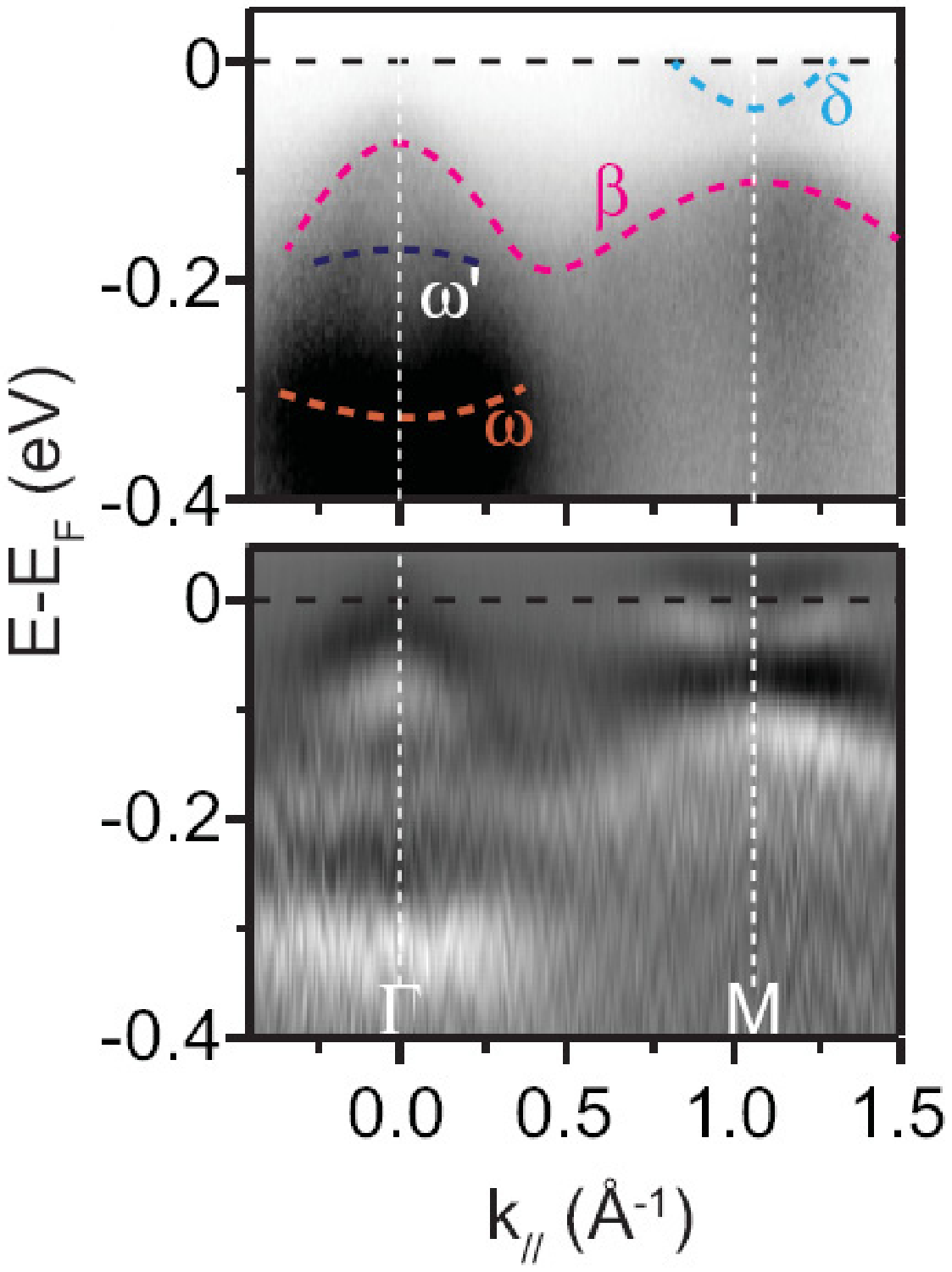}
\caption{(a) -- LDA bands for LiOHFeSe
(Fermi level is at $E=$0) \cite{LiOHFeSe_NS},
(b) -- LDA Fermi surface LiOHFeSe, corresponding to electron doping of
0.3 electrons per unit cell,
(c) -- experimental ARPES Fermi surfaces for [Li$_{0.8}$Fe$_{0.2}$OH]FeSe
\cite{LiOHFeSe_ARP},
(d) -- experimental ARPES bands near the Fermi level of
[Li$_{0.8}$Fe$_{0.2}$OH]FeSe \cite{LiOHFeSe_ARP}.}
\label{LiOHFeSe_spectr_FS}
\end{figure}

In Ref. \cite{LiOHFeSe_NS} we presented the results of LDA calculations for
stoichiometric LiOHFeSe. Corresponding energy band dispersions are shown in
Fig. \ref{LiOHFeSe_spectr_FS} (a). At first glance, the energy spectrum of this
system is quite similar to the spectra of the most of FeAs based systems and
bulk FeSe. In particular, the main contribution to the density of states in the
wide energy range around the Fermi level is determined solely by Fe-3$d$.
The Fermi surfaces are qualitatively similar to that of majority of Fe
based superconductors. However, this is somewhat misleading impression.
Real [Li$_{0.8}$Fe$_{0.2}$OH]FeSe system, where superconductivity was
discovered, the partial replacement of Li by Fe in LiOH intercalation layers
leads to noticeable LiOH {\em electron} doping, so that the Fermi energy moves
upwards (relative to the stoichiometric case) by about 0.15 - 0.2 eV.
Then, as can be seen from Fig. \ref{LiOHFeSe_spectr_FS} (a),
the hole bands close to the $\Gamma$ -- point shifts below the Fermi level
and the hole cylinders of Fermi surface almost disappear.
The general view of the LDA calculated Fermi surface for this level of electron
doping is shown in Fig. \ref{LiOHFeSe_spectr_FS} (b).
It has an obvious similarity with the results for the A$_x$Fe$_{2-y}$Se$_2$
system (see Fig. \ref {FSAFeSe}).
These conclusions are directly confirmed by ARPES experiments \cite{LiOHFeSe_ARP},
with the results are shown in Fig. \ref{LiOHFeSe_spectr_FS} (c).

From Fig. \ref{LiOHFeSe_spectr_FS} one can see that the Fermi surface consists
mostly of electronic cylinders surrounding the $M$-points, while around the
$\Gamma$ -- point the Fermi surface is either absent or very small.
In any case, for this system we can not speak of any ``nesting'' of electron and
hole Fermi surfaces in any sense.  Electronic dispersions found in the ARPES
experiments are very similar to corresponding ARPES dispersions reported in Refs.
\cite{KFe2Se2_ARPES, KFe2Se2_ARPES_2} for K$_{1-x}$Fe$_{2-y}$Se$_2$ system.
These are qualitatively similar to dispersions obtained in LDA and LDA+DMFT
calculations, including rather strong correlation bands narrowing
(by about several times with different compression factor for different bands)
\cite{KFeSeLDADMFT1, KFeSeLDADMFT2}).
However, the explanation of the formation of extremely ``shallow'' electron
$\delta$ band with depth $\sim$0.05 eV near the $M$-point remains unclear.
This requires an unusually strong correlation compression which hardly can be
obtained from the LDA+DMFT calculations, while the diameter of electronic
cylinders around $M$ -- points is nearly unchanged by correlations and almost
coincides with the results of LDA calculations.

An interesting debate flared up around the possible nature of magnetic ordering
of Fe ions, which replaces Li ions within intercalation LiOH layers.
In Ref. \cite{LiOH1} it was stated that this ordering is just a canted
antiferromagnet. However, in Ref. \cite{LiOH2}, on the basis of magnetic
measurements, it was claimed that it is  ferromagnetic, with Curie temperature
$T_C\sim$10K, i.e. substantially below the superconducting transition temperature.
This conclusion was confirmed indirectly in Ref. \cite{LiOH3} by observing the
scattering of neutrons on the lattice of Abrikosov vortices, which might be
induced in the FeSe layers by ferromagnetic ordering of Fe spins in the
Li$_{1-x}$Fe$_x$OH layers. At the same time, in Ref. \cite{LiOH4} it was argued
that M\"ossbauer measurements indicate the absence of any kind of magnetic
ordering on Fe ions in intercalation layers.

In Ref. \cite{LiOHFeSe_NS} LSDA calculations of the exchange integrals were
performed for some typical magnetic configurations of Fe ions, replacing Li in
LiOH layers. For the most likely magnetic configuration, leading to magnetic
ordering, we have obtained the positive (ferromagnetic) sign of the exchange
interaction, and simple estimate of the Curie temperature produced the value of
Curie temperature $ T_C \approx $ 10K, which is in excellent agreement with the
results of experiment of Refs. \cite{LiOH2, LiOH3}.

\subsection{FeSe monolayer films}

\begin{figure}
\includegraphics[clip=true,width=0.45\textwidth]{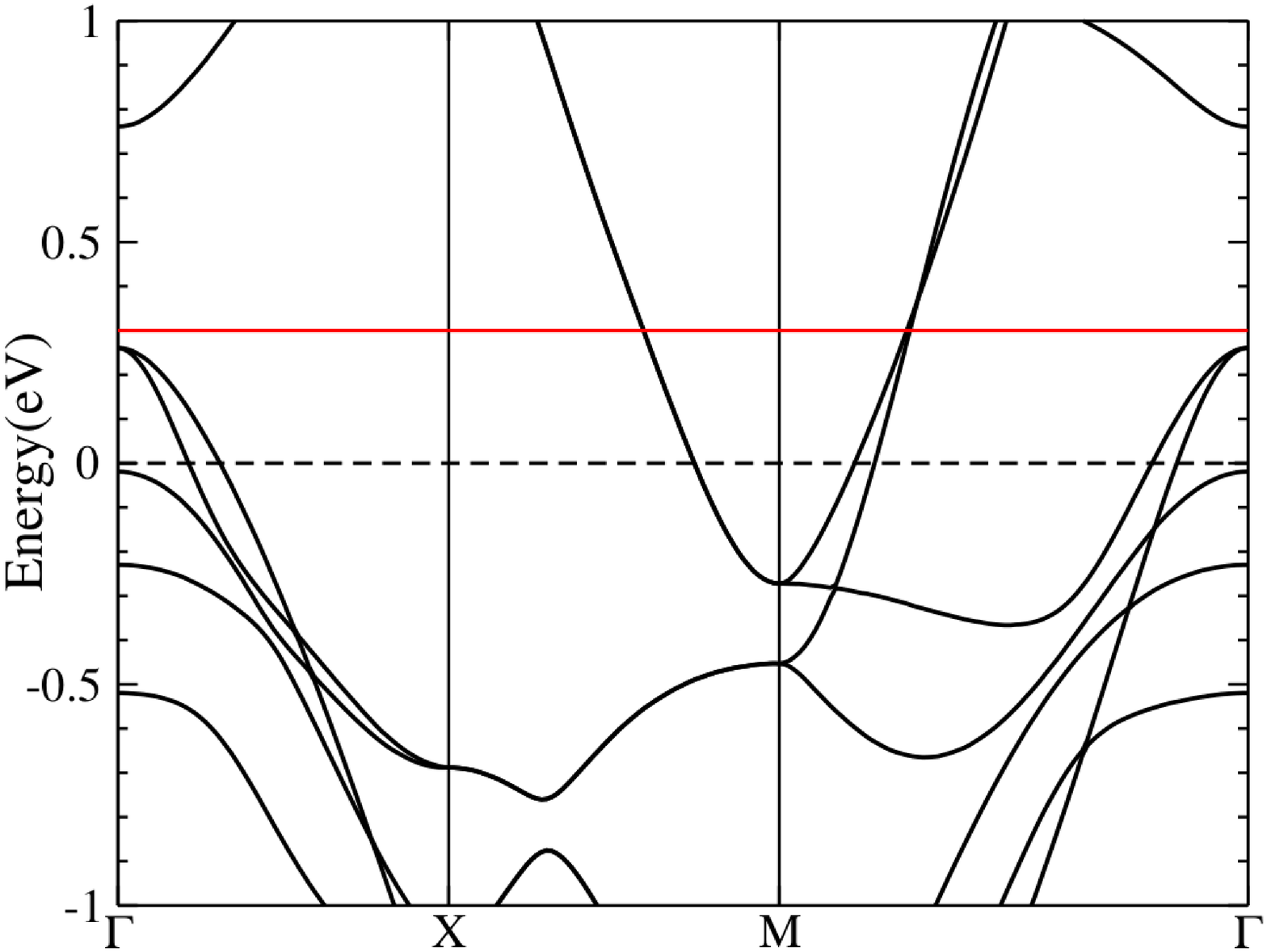}
\includegraphics[clip=true,width=0.5\textwidth]{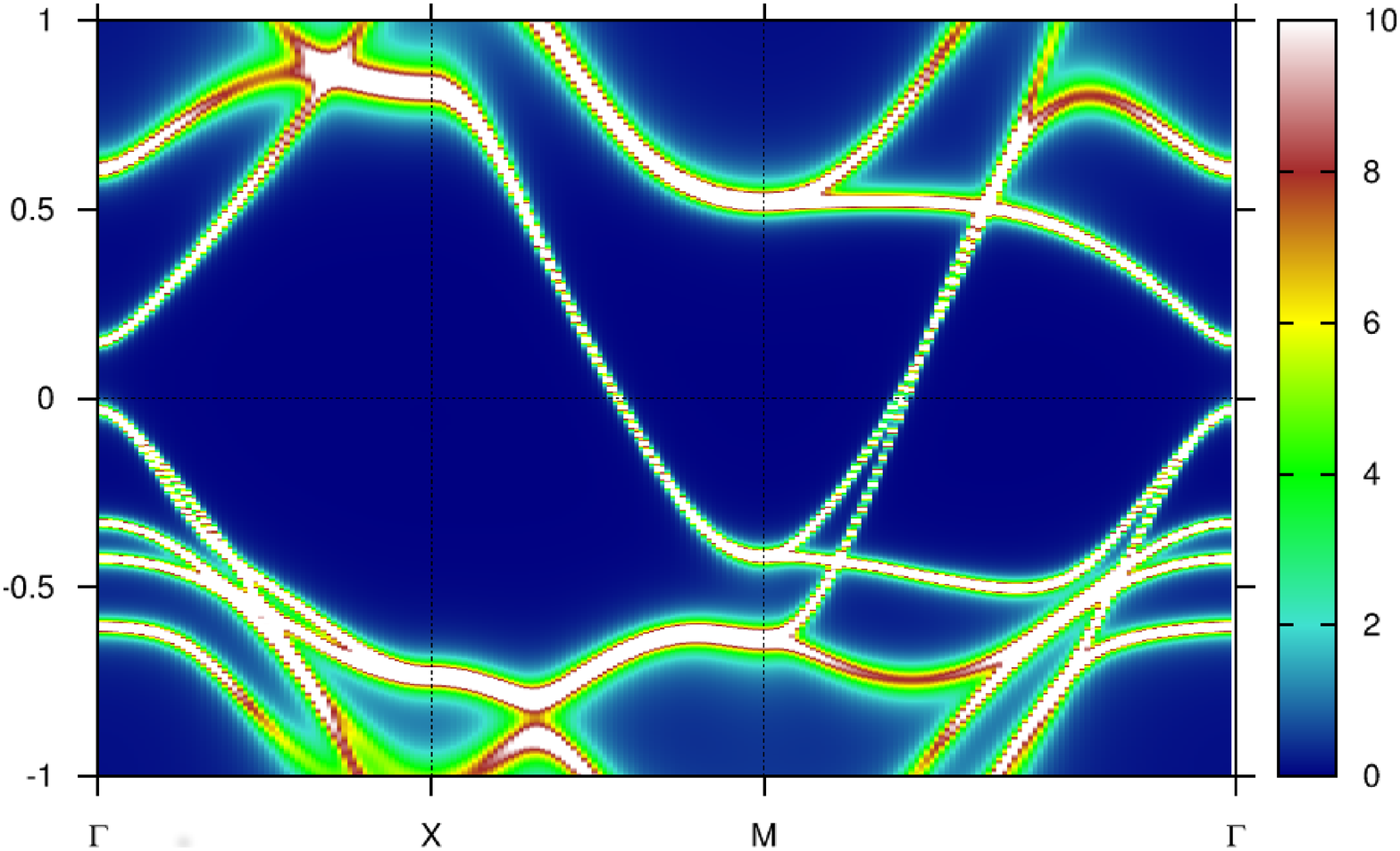}
\caption{(a) -- LDA bands of an isolated FeSe monolayer near the Fermi level
($E=$0). Red line shows approximate Fermi level position to agree with ARPES
experimental data.
(b) -- LDA+DMFT bands of isolated electronically doped FeSe monolayer near
the doping -- shifted Fermi level.}
\label{FeSe_LDA_DMFT_bands}
\end{figure}

\begin{figure}
\includegraphics[clip=true,width=0.45\textwidth]{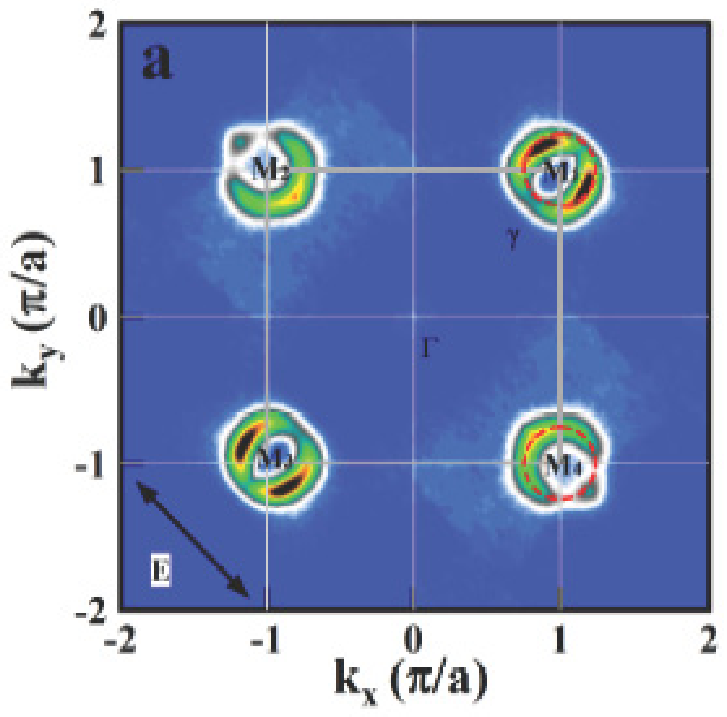}
\includegraphics[clip=true,width=0.5\textwidth]{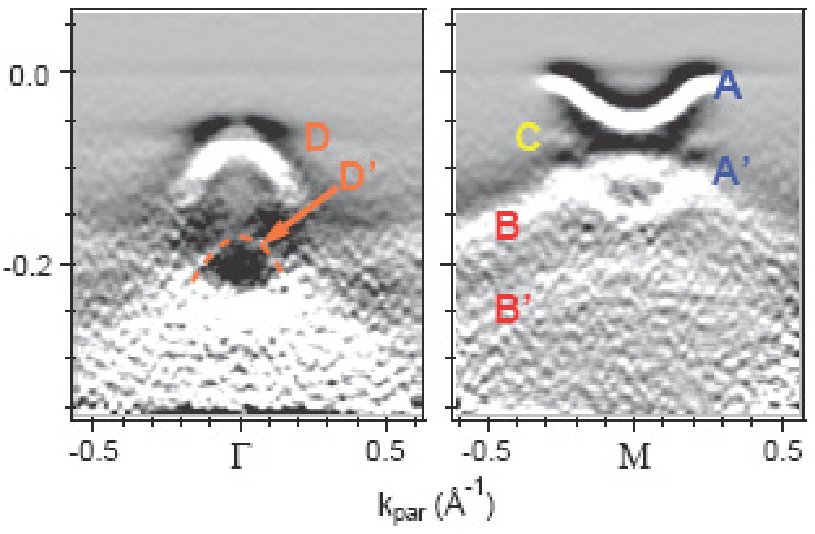}
\caption{(a) -- Experimental ARPES Fermi surface of FeSe monolayer
\cite{ARP_FS_FeSe_1},
(b) -- Experimental ARPES bands of FeSe monolayer near the Fermi level
\cite{ARPES_FeSe_Nature}.}
\label{FeSe_ARPES_bands}
\end{figure}

LDA calculations of the spectrum of the isolated FeSe monolayer can be performed
in a standard slab approach. To calculate electronic properties we used the
{\em Quantum-Espresso} \cite{qe} package. The results of these calculations are
shown in Fig. \ref{FeSe_LDA_DMFT_bands} (a).
It can be seen that the spectrum has the form typical for FeAs based systems and
bulk FeSe discussed in detail above.
However ARPES experiments \cite{ARP_FS_FeSe_1, ARPES_FeSe_Nature, ARP_FS_FeSe_2}
convincingly show that this is not so. For FeSe monolayers on STO only
electronic Fermi surface sheets are observed around the $M$ -- points of the
Brillouin zone, while hole sheets, centered around the $\Gamma$ -- point (in the
center of the zone), are simply absent. An example of such data is shown in
Fig. \ref{FeSe_ARPES_bands} (a) \cite{ARP_FS_FeSe_1}.
Similarly to intercalated FeSe systems there are no signs of ``nesting'' of
Fermi surface -- there are just no surfaces to ``nest''!

In an attempt to explain the contradiction between ARPES experiments
\cite{ARP_FS_FeSe_1} and band structure calculations reflected in the
absence of hole cylinder in the $\Gamma$ -- point, one can suppose that this may
be the consequence of FeSe/STO monolayer stretching due to mismatch of lattice
constants of the bulk FeSe and STO. We have studied this problem by varying the
lattice parameter $a$  and Se height $z_{Se}$ in the range $\pm 5\%$ around the
bulk FeSe parameters. Before the electronic structure was calculated crystal
structure was relaxed. Unfortunately, the conclusion was that the changes of
lattice parameters do not lead to qualitative changes of FeSe Fermi surfaces and
the hole cylinders in the $\Gamma$ -- point always remain.

However, there is another rather simple qualitative explanation for the absence
of hole cylinders and the observed Fermi surfaces can be obtained by assuming
that the system is just electronically doped.  The Fermi level has to be moved
upwards in energy by the value of $\sim $ 0.2 - 0.25 eV, as shown
by the solid (red) horizontal line in Fig. \ref{FeSe_LDA_DMFT_bands} (a),
which corresponds to the doping level of 0.15 - 0.2 electron per Fe ion.

Strictly speaking, the nature of this doping is not fully identified. But there
is a common belief that it is associated with the formation of oxygen vacancies
in the SrTiO$_3$ substrate (within the topmost layer of TiO$_2$),
occurring during the various technological steps used during the film preparation,
such as annealing, etching, etc. It should be noted that the formation of the
electron gas at the interface with the SrTiO$_3$ is a widely known phenomenon,
which was studied for a long time \cite {Shkl_16}. At the same time, for
FeSe/STO system this issue remains poorly understood (see, however, Refs.
\cite{Mills, Agter}.

Electron correlations have relatively little influence on the spectrum of the
FeSe monolayer. Fig. \ref{FeSe_LDA_DMFT_bands} (b) shows the results of LDA+DMFT
calculations for estimated experimental value of electron doping.
DMFT calculations were done for the values of the Coulomb and exchange (Hund)
electron interactions in the Fe-3d shell equal to $ U = 3.5 $ ~ eV and
$ J = 0.85 $ eV. As {\em impurity solver}  we used the Continuous Time Quantum
Monte -- Carlo algorithm (CT-QMC). The inverse dimensionless temperature was
taken to be $\beta$ = 40. It can be seen that the spectrum is very weakly
renormalized by correlations and preserves the general LDA -- like shape,
with a slight band compression factor of about $\sim $ 1.3.

\begin{figure}[h]
\begin{minipage}[h]{0.62\linewidth}
\center{\includegraphics[width=0.95\linewidth]{FeSe_SrTiO3.eps} }
\end{minipage}
\hfill
\begin{minipage}[h]{0.65\linewidth}
\center{\includegraphics[width=0.95\linewidth]{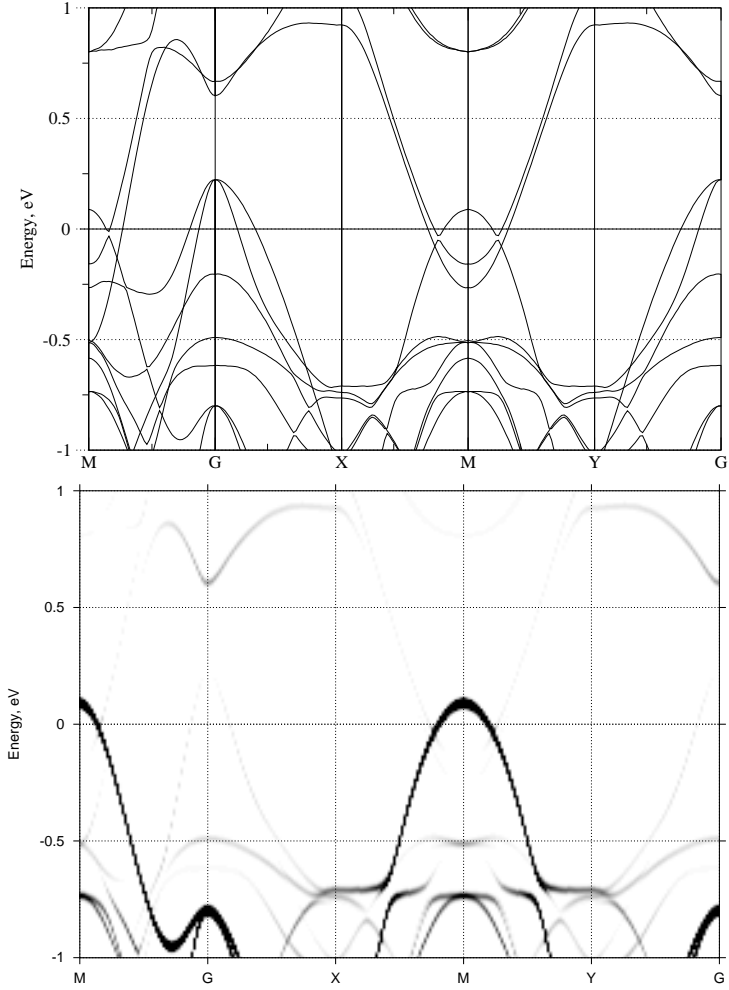} }
\end{minipage}
\caption{(a) -- LDA calculated band dispersion of FeSe monolayer on
SrTiO$_3$ substrate;  (b) -- LDA calculated O-2p states of surface TiO$_2$
layer of STO substrate.}
\label{mlFeSe_STO}
\end{figure}

Electronic dispersion in monolayer FeSe films was measured by ARPES and reported
in a number of papers \cite{FeSe_BTO, ARPES_FeSe_Nature}. The results of Ref.
\cite{ARPES_FeSe_Nature} are shown in Fig. \ref{FeSe_ARPES_bands} (b). They are
consistent with those of other studies and, in general, are similar to those
obtained for the intercalated FeSe systems
(see e.g. Fig. \ref{LiOHFeSe_spectr_FS} (c)).
Overall, they are qualitatively similar to the results of LDA+DMFT but there is
no quantitative agreement. In particular, in the ARPES experiments the presence
of the unusual ``shallow'' electron band at $M$ -- point, with the band bottom
at $\sim $ 0.05 eV is clearly observed. However, our calculated dispersions
show a ``depth'' of almost an order of magnitude larger.

It should also be noted that an additional ``shadow'' or ``replica'' electronic
band near the $M$ -- point was observed in Ref. \cite{ARPES_FeSe_Nature}, which
is $\sim$100 meV below the parent band ``shallow'' band, and is clearly visible
in Fig. \ref{FeSe_ARPES_bands} (b). This ``shadow'' band
is completely missed in the band structure calculations. The possible nature of
this band (due to interaction with 100 meV STO optical phonons)
and its importance for mechanisms of $T_c $ enhancement in FeSe/STO films was
discussed in Ref. \cite{ARPES_FeSe_Nature}.

Now let us discuss the results of our LDA calculations of electronic structure
of the FeSe monolayer film on SrTiO$_3$ substrate as shown in Fig. \ref{FeSeSTO}.
These calculations were again performed with {\em Quantum Espresso} \cite{qe}.
By looking on left panel of Fig. \ref{mlFeSe_STO} one can immediately recognize
that there appears an additional hole band near the $M$ -- point. To understand
its origin we plotted on right panel of Fig. \ref{mlFeSe_STO} O-2p states of
topmost surface TiO$_2$ layer of STO substrate. We can conclude, that the
presence of STO interface leads to the appearance of this additional band
of O-2p surface states near the Fermi level with probable ``nesting'' with hole
Fe 3d -- band band with nesting vector ${\bf Q}$=0. Also we observe rather small
splitting of electron bands at $M$ -- point. The relevance of these results to
ARPES experiments and their significance for enhanced superconductivity in
FeSe/STO system are at present unclear. However, the additional electron band
splitting may have the relation to the observation of the ``replica'' band at
$M$ -- point \cite{ARPES_FeSe_Nature}, providing the explanation of its
appearance not related to interaction with optical phonons in STO.

\section{Conclusion}

We have presented a short review of calculations of electronic band
structure of high -- $T_c$ systems based on FeSe monolayers, from intercalated
compounds like A$_x$Fe$_2$Se$_{2-z}$S$_z$ (A=K,Cs,...) and
[Li$_{1-x}$Fe$_x$OH]FeSe to single FeSe layer films of SrTiO$_3$ substrate.

Our calculations show, that in all these systems the general structure of
electronic spectrum is significantly different from the ``standard model''
typical for almost all FeAs based superconductors and bulk FeSe. This structure
is characterized by the practical absence of hole -- like Fermi surface
cylinders around the $\Gamma$ -- point at the center of the Brillouin zone with
only electron -- like cylinders present around the $M$ -- points in the corners
of the Brillouin zone. These results are essentially confirmed by the available
ARPES experiments. An apparent absence of obvious ``nesting'' properties
between electron and hole Fermi surface pockets cast serious doubts upon the
most popular picture of Cooper pairing, assumed to realize in FeAs systems,
based on the exchange of antiferromagnetic fluctuations and leading to the
picture of $s^{\pm}$ pairing \cite{MazKor}.

The role of electronic correlations in FeSe monolayers remain rather
controversial. LDA+DMFT calculations for K$_x$Fe$_2$Se$_2$S system
show that here these correlations are more important than in typical FeAs
systems, leading to stronger bandwidth compression (different for different
Fe 3d -- bands) and rather poorly defined quasiparticles close to the Fermi
level. In contrast to this, in isolated FeSe single -- layer system the same
calculations demonstrate rather weak influence of correlations with small
bandwidth compression (effective mass renormalization).  The role of SrTiO$_3$
substrate may also be important leading to the appearance of O-2p hole band
of TiO$_2$ layer in the vicinity of the Fermi level.

The serious problems for understanding of the observable electronic structure of
FeSe monolayer systems remain to be solved. In particular, at present there are
no acceptable explanation of formation of unusually ``shallow'' electronic
bands around the $M(X)$ -- points, with Fermi energies $\sim$ 0.05 eV, observed in
ARPES experiments on all of these systems. It can be guessed that this fact
reflects our poor understanding of electron correlations and the
question remains open. Note that the existence of such small Fermi energies in
conduction bands of the system under discussion signifies the serious problems
related to the antiadiabatic regime of Cooper pairing \cite{Gork_1}. Taking
into account the typical values of the energy gap in FeSe monolayer superconductors
$\Delta\sim$ 0.015-0.02 eV \cite{FeSe_BTO,ARP_FS_FeSe_2} we also obtain the
anomalously low gap to the Fermi energy ratios: $\Delta/E_F\sim$0.25-0.5, which
indicate, that these superconductors belong to BCS -- BEC crossover region
\cite{NozSR,Rand}.

\section{Acknowledgments}

Calculations of electronic structure of FeSe  systems 
were performed under the FASO state contract No. 0389-2014-0001 and
partially supported by RFBR grant 14-02-00065. NSP and AAS work was 
also supported by the President of Russia grant for young scientists
No. Mk-5957.2016.2.


\end{document}